# Breaking the energy-bandwidth limit of electro-optic modulators: theory and a device proposal

Hongtao Lin, *Student Member, IEEE*, Okechukwu Ogbuu, Jifeng Liu, Lin Zhang, Jurgen Michel, and Juejun Hu, *Member, IEEE*

*Abstract*—In this paper, we quantitatively analyzed the trade-off between energy per bit for switching and modulation bandwidth of classical electro-optic modulators. A formally simple energy-bandwidth limit (Eq. 10) is derived for electro-optic modulators based on intra-cavity index modulation. To overcome this limit, we propose a dual cavity modulator device which uses a coupling modulation scheme operating at high bandwidth (> 200 GHz) not limited by cavity photon lifetime and simultaneously features an ultra-low switching energy of 0.26 aJ, representing over three orders of magnitude energy consumption reduction compared to state-of-the-art electro-optic modulators.

*Index Terms*— Electrooptic modulators, photonic crystals, optical switches, optical resonators

## I. INTRODUCTION

Optical modulators and switches are critical building blocks for optical communication networks. As optical links progressively evolve to replace electrical wirings at the board and chip levels, power consumption and speed are increasingly becoming the limiting factors to further scale down the length of optical interconnects. To compete with existing electrical interconnect technologies, the energy consumption of optical devices has to be in the order of 10 fJ/bit or lower according to an analysis presented by Miller [1].This low energy consumption poses a significant design challenge for optical modulators: traditional LiNbO$_3$ Mach-Zehnder (M-Z) interferometer modulators typically operate at ~ 10 pJ/bit or above. State-of-the-art silicon-based electro-absorption and electro-optic modulators currently achieve a typical switching energy of 3 - 100 fJ/bit [2-10]. Switching energy down to < 10 fJ/bit was also attained in plasmonic nanostructures with strong optical confinement at the cost of increased insertion loss due to metal absorption [11-13].

One solution to the aforementioned challenge is to leverage optical resonant enhancement effects to reduce the footprint and thus energy consumption of modulator devices. In principle, the ultimate low switching energy can be attained in a nanoscale cavity which supports only a single resonant mode. Besides size scaling, cavities with high quality factors (Q) are extremely sensitive to intra-cavity refractive index perturbations or absorption change. Due to this unique feature, increasing cavity Q leads to reduced switching energy. Nevertheless, high-Q also results in long cavity photon lifetime which limits the modulation bandwidth. Performance constraints and scaling laws in resonant optical modulators have been published [14-16], although the numbers quoted were based on specific material choices (e.g. Si or LiNbO$_3$) and device geometries. It is not yet clear what is the ultimate fundamental limit to modulation energy scaling. In addition, the trade-off between switching energy and modulation bandwidth has not been quantitatively analyzed.

In the first part of this paper we present a full quantitative analysis on the fundamental energy-bandwidth limit of classical Electro-Optic (EO) intra-cavity modulated resonant modulator devices. We choose to investigate modulators based on the Pockels effect, since this mechanism does not involve physical displacement of charged carriers and thus has an inherent temporal response down to the femtosecond range which enables ultrafast modulation beyond 1 THz. The theoretical formalism presented here, however, is also applicable to electro-optic modulators based on free carrier plasma dispersion, and can be extended to electro-absorption modulators in a straightforward manner. Further, even though this theoretical limit is derived for resonant modulator devices, the performance of non-resonant modulators (e.g. M-Z interferometers) is bound by this limit as well.

In the second part of the paper we propose a coupling modulated dual cavity design to overcome the intra-cavity EO modulation limit. The concept of coupling modulation, previously analyzed in micro-rings by several authors[17-20], enables modulation rates approaching the micro-ring's free spectral rang (FSR) in theory, which far exceeds the cavity photon lifetime limit. The coupling modulation approach has also been successfully implemented in experiment to

Manuscript received XXX. This work was supported by the National Science Foundation under award number 1200406 and a Delaware EPSCoR Seedling grant.
Hongtao Lin, Okechukwu Ogbuu and Juejun Hu are with the University of Delaware, Newark, DE 19716, USA (e-mail: hujuejun@udel.edu).
Jifeng Liu is with Thayer School of Engineering, Dartmouth College, Hanover, NH 03755, USA
Lin Zhang and Jurgen Michel are with the Massachusetts Institute of Technology, Cambridge, MA 02139, USA.

dynamically control the extrinsic Q-factor of a photonic crystal cavity [21]. By incorporating a novel dual photonic crystal (PhC) / micro-ring resonant cavity structure, we show that the device can exhibit fast modulation exceeding 200 GHz bandwidth while maintaining an ultra-low switching energy of in the order of 0.1 aJ, which is more than three orders of magnitude lower compared to state-of-the-art electro-optic modulator devices.

## II. THE ENERGY-BANDWIDTH LIMIT OF INTRA-CAVITY MODULATED EO RESONATORS

Here we would like to derive the theoretical performance limits of resonant modulators rather than a specific class of device architecture (e.g., micro-rings or photonic crystal cavities); therefore, we examine a generic resonant cavity structure coupled to an input and an output light beam as shown in Fig. 1a: the cavity refractive index is electro-optically modulated by an applied RF electric field, resulting in spectral shift of the resonant peak and change of transmitted light intensity (Fig. 1b). We first assume the cavity is critically coupled, as critical coupled cavity gives minimal insertion loss and maximum extinction ratio. This condition means that the cavity is lossless, and the input and output coupling strengths are equal:

$$\tau_{in} = \tau_{out} = \tau \quad (1)$$

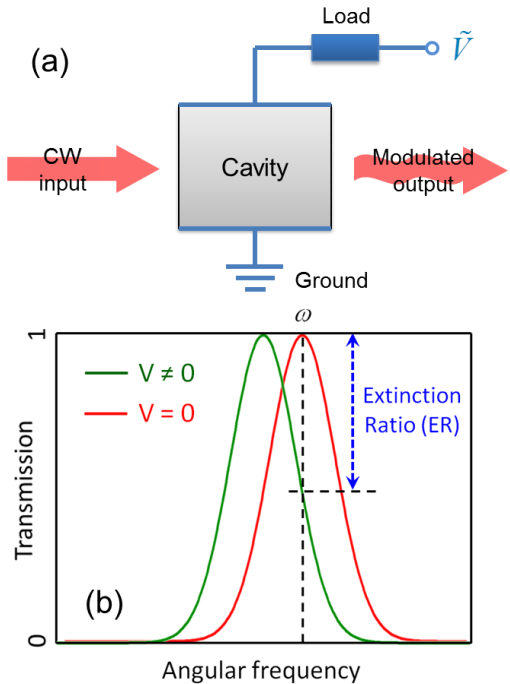

Fig. 1. (a) Schematic resonant modulator device layout, showing a resonant cavity whose refractive index is modulated by electro-optic effects; (b) illustration of the electro-optic modulation mechanism: resonant frequency of the cavity is modulated by the intra-cavity index change, resulting in transmitted intensity modulation

where $\tau_{in}$ and $\tau_{out}$ denote the input and output coupling limited cavity photon lifetime, respectively. Note that even though Fig. 1a shows a transmissive filter type cavity, our theory is equally applicable to reflective or all-pass filter types of cavity as well: the only modification necessary is to replace the output coupling limited lifetime with intrinsic photon lifetime in the critical coupling condition Eq. 1.

In a high speed electro-optic modulator, the main switching energy-per-bit contributor is the power generator resistive energy dissipation during the charge-discharge cycle of the capacitor, given by:

$$W = \frac{1}{2} C V_{pp}^{2} \quad (2)$$

where $C$ is the capacitance of the cavity and $V_{pp}$ is the peak-to-peak full-swing driving voltage. Note that the result is independent of the generator resistance. According to Miller [22], the electrical energy required for charging a capacitor, which is also given by $\frac{1}{2}CV_{pp}^2$, can be recovered by using a by-pass capacitor. This formula gives the upper limit of switching energy per bit irrespective of data coding scheme; for a Non-Return-to-Zero (NRZ) sequence with roughly equal bits of 1's and 0's, the average energy per bit will only be half of the value given by Eq. 2. Since the energy per bit is independent of the capacitor shape to the first order (refer to Appendix I for proof), we assume a cubic shape cavity with edge length $L$ which supports only a single resonant mode. Such a single mode cavity offers minimal capacitance and switching energy. To effectively confine one optical mode in the cavity at the resonant wavelength $\lambda$, the edge length $L$ can be derived from the wave vector quantization condition:

$$k_x^{2} + k_y^{2} + k_z^{2} = n^2 k_0^{2} \quad (3)$$

which yields:

$$L = \frac{\sqrt{3}}{2} \cdot \frac{\lambda}{n} \quad (4)$$

Here $n$ is the refractive index of the cavity material. This condition holds for non-metals whose permittivity is positive. Eq. 4 gives the mode volume required to completely confine an optical mode (i.e. unity optical confinement factor) in a dielectric cavity. The minimum capacitance of the cubic cavity is thus given by:

$$C = \Gamma \varepsilon \frac{L^2}{L} = \frac{\sqrt{3}}{2} \cdot \frac{\Gamma \varepsilon \lambda}{n} \quad (5)$$

where $\varepsilon$ represents the permittivity of the cavity material at the modulation frequency and $\Gamma$ is a geometric factor of the order of unity. Without losing generality, henceforth we take $\Gamma = 1$. At the telecommunication wavelength $\lambda = 1.55$ μm, the RC delay time for such a capacitor with 50 ohm resistive load is ~ 1 fs. Therefore, cavity photon lifetime (~ ps), rather than the exceedingly small RC time constant, poses the main constraint on high-speed modulation in the diffraction-limited cavity modulator. Further, since the Pockels effect is electronic in nature and has a very short inherent response time (~ fs), the cavity material EO refractive index change instantaneously follows the applied electric field $V(t)$. Therefore, the resonant frequency shift of the cavity $\Delta\omega$ also exhibits an instantaneous response to $V(t)$:

$$\Delta\omega = \frac{\Delta n_{eff}}{n_g} \cdot \omega \sim \frac{n^2 r \omega V}{2L} = \frac{n^3 r \omega^2}{2\sqrt{3}\pi c_0} \cdot V \quad (6)$$

Here $c_0$ is the speed of light in vacuum, $r$ is the linear electro-optic coefficient of the cavity material [23], $\omega$ is the cavity resonant frequency, and we make the valid approximation that the electric field in the cavity $E \sim V/L$. Note that in Eq. 6 the '~' symbol was given by taking the group index to be equal to the effective index; this is certainly not necessarily the case in slow light cavities, where the ratio $n_{eff}/n_g$ can be much smaller than unity. However, the change is exactly cancelled out by a confinement factor increase (note that the confinement factor can be greater than unity in a slow light cavity) and thus modulator devices utilizing slow light cavities are also bound by the limit we propose in this paper.

To analyze the modulation frequency dependence of cavity response, we resort to the Coupled Mode Theory (CMT) [24], Previous studies indicate that the CMT model, which treats the cavity as a lumped circuit element, generates almost identical results to Sacher's approach [17] when applied to single-cavity modulators [25, 26]. According to the CMT model, time-dependent optical transmittance of the cavity shown in Fig. 1a in response to a sinusoidally varying voltage $V(t) = -\frac{1}{2}V_{pp} \cdot \sin(\Omega t)$ is:

$$T(t) = \frac{2}{\tau} \cdot \exp\left(\frac{i\Delta\omega_{pp}t}{2} - \frac{2t}{\tau} + \frac{i\Delta\omega_{pp}\cos\Omega t}{2\Omega}\right)$$
$$\cdot \int_0^t \exp\left(-\frac{i\Delta\omega_{pp}t'}{2} + \frac{2t'}{\tau} - \frac{i\Delta\omega_{pp}\cos\Omega t'}{2\Omega}\right) \cdot dt' \quad (7)$$

where $\Delta\omega_{pp}$ gives the peak resonance frequency shift as:

$$\Delta\omega_{pp} = \frac{n^3 r \omega^2}{2\sqrt{3}\pi c_0} \cdot V_{pp} \quad (8)$$

and $\tau$ is the input (or output) coupling limited cavity photon lifetime defined in Eq. 1. Derivation of Eq. 7 using the CMT formalism is elaborated in Appendix II. An example of the time-domain output given by Eq. 7 is plotted in Fig. 6b. Extinction ratio (i.e. modulation depth) in dB at the modulation frequency $f = \Omega/2\pi$ is thus trivially defined as:

$$ER = 10 \cdot \log_{10}\left(\frac{T_{max}}{T_{min}}\right) \quad (9)$$

Combining Eqs. 2, 5, 7 and 9, the energy per bit $W$ can be expressed as a function of the modulation frequency $f = \Omega/2\pi$ and the modulation extinction ratio. Here we define $f_{3dB}$ as the optical 3 dB bandwidth when the extinction ratio rolls off to 3 dB at a given $V_{pp}$ [27]. As an example, we assume a cavity material with a refractive index $n = 1.5$ and a linear EO coefficient $r = 300$ pm/V (both are typical values in highly nonlinear EO polymers [28]), and the switching energy per bit is plotted as a function of $f_{3dB}$ in Fig. 2 in cavities operating at 1550 nm wavelength, with different Q-factors. High-Q cavities exhibit spectrally sharp resonant peaks and thus require smaller electro-optic frequency shift to reach 3 dB extinction ratio at low modulation frequencies; however, their long cavity photon lifetime leads to lower onset frequency of the switching energy rise. The yellow region in Fig. 2 corresponds to the performance domain attainable using classical intra-cavity modulated devices. The minimum energy per bit $W_{lim}$ of classical EO modulators can be derived by fitting the limiting behavior of the curves at their high frequency ends, which leads to:

$$W_{lim} = K \cdot f_{3dB}^2 \quad (10)$$

Here $K$ is a material constant given by $K = 0.6\varepsilon_r n^{-7} r^{-2}$ fJ/(GHz·pm/V)$^2$ at 1550 nm wavelength, where $\varepsilon_r$ is the relative permittivity at the modulation frequency and the material EO coefficient $r$ is given in pm/V. This exceedingly simple result places a lower limit to the switching energy in classical EO resonant modulators. Since $K$ describes the electro-optic behavior of the active material, Eq. 10 is general and **not** material-specific. At a modulation frequency of 10 GHz, this limit suggests a minimal energy per bit as low as ~0.1 aJ, which is several order of magnitude lower than that of state-of-the-art EO modulators. However, due to the quadratic dependence of switching energy on modulation frequency, the power penalty rapidly increases at high frequencies, posing a major hurdle for ultrafast optical modulation.

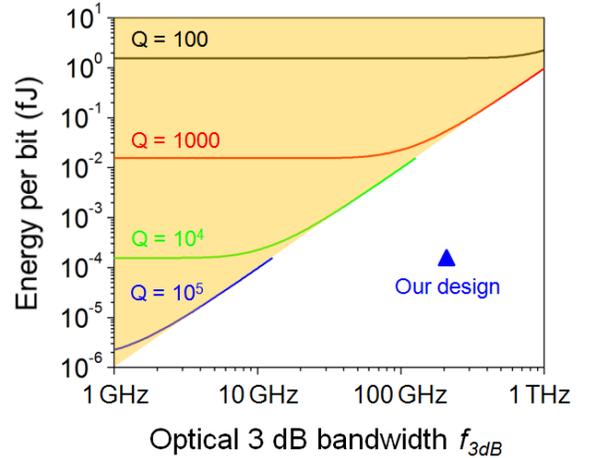

Fig. 2. Trade-off between switching energy and modulation frequency at a fixed extinction ratio of 3 dB in classical electro-optic modulators (assuming a 300 pm/V EO coefficient). Ultimate theoretical performance of resonant modulators with different Q-factors (curves) and performance domains attainable in classical EO modulators (yellow) are depicted. Performance of the proposed dual cavity modulator is also plotted for comparison.

### III. APPLICABILITY OF THE MODEL

According to Koch [16], the energy consumption of resonant modulators is roughly reduced by a factor equaling cavity finesse compared to their non-resonant counterparts. Therefore, while the theoretical formalism in the preceding section is derived for resonant modulators, the performance of non-resonant electro-optic modulators (e.g. M-Z interferometers) is bound by this limit as well.

The switching energy limit also applies to modulators based on free carrier plasma dispersion effect. For example, for a modulator operating in a carrier injection mode, the refractive index change $\Delta n = n^3 Er/2$ in Eq. 6 should be replaced by:

$$\Delta n = (K_n + K_p) \cdot \exp\left(\frac{eV}{2k_B T}\right) \quad (11)$$

where $K_n$ and $K_p$ represent the linearized index change coefficients, $k_B$ is the Boltzmann constant, and $T$ is the junction temperature. If the device bandwidth is limited by cavity photon lifetime rather than carrier dynamics, the remainder of the theoretical framework still follows an identical formalism.

Lastly, in Eq. 3, if at least one of the wave vector components is allowed to be imaginary (for instance, in the case of a plasmonic nano-cavity), it is possible to achieve deep sub-$\lambda$ 3-D confinement in a single-mode cavity and hence even smaller capacitance than the value given by Eq. 5. The parasitic optical loss due to metal absorption and coupling loss to the sub-wavelength nano-cavity, however, need to be mitigated by careful design optimization.

## IV. COUPLING MODULATED DUAL CAVITY MODULATOR DESIGN: BEYOND THE CLASSICAL LIMIT

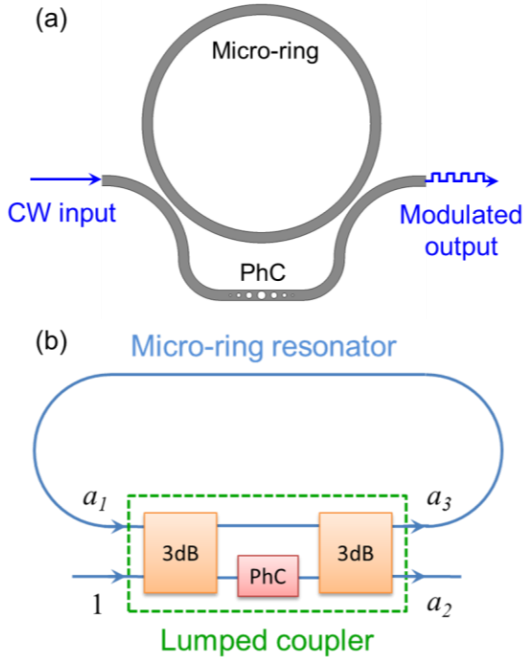

Fig. 3. (a) Schematic design of the dual photonic crystal/micro-ring cavity modulator (top view); (b) block diagram of the dual cavity modulator: coupling to the ring is modulated by tuning the phase of light passing through PhC cavity.

While the bandwidth of intra-cavity modulation is fundamentally bound by cavity photon lifetime, it has been shown that this limit can be circumvented by adopting a coupling modulation scheme [17, 20]. Here we show that by combining the coupling modulation scheme with a novel dual cavity design, it is possible to break the theoretical performance limit of classical EO modulators. Figure 3 shows the schematic design of the device: it consists of a micro-ring resonator incorporating an M-Z interferometer coupler; a horizontal slot waveguide nanobeam PhC cavity filled with EO polymers is embedded in one of the interferometer arms to electro-optically control the relative phase delay between the two arms. The phase delay is set to be $\pi$ when the applied voltage is zero, which leads to destructive interference and zero coupling into the cavity, and hence near unity transmission in the through port corresponding to the "on" state. When voltage is applied across the PhC cavity, the EO index change leads to a phase shift which enables coupling into the ring. Unlike conventional MZ interferometers which require a $\pi$ phase shift to switch between the "on" and "off" states, full-swing phase shift (and voltage) of the dual cavity modulator is $\ll \pi$, since the high-Q micro-ring is very sensitive to coupling perturbation. On the other hand, speed of the modulator device is limited by two factors, the micro-ring FSR and the bandwidth of the PhC cavity. Ultra-high speed operation (from a few hundred GHz up to 1 THz) is attainable if a relatively low-Q (~1,000 or below) PhC cavity is used. In this way, by integrating a high-Q micro-ring with a low-Q PhC cavity, the dual cavity design combines the merit of high Q cavities in low power switching with the merit of low Q cavities in short photon lifetime.

As an added benefit, the dual cavity structure also enables independent optimization of the device's electro-optic and athermal performance by applying different coatings on the micro-ring and the PhC cavity. For example, robust post-fabrication trimming [29] and athermal operation [30, 31] may be simultaneously achieved by cladding the micro-ring with a polymer-glass bi-layer.

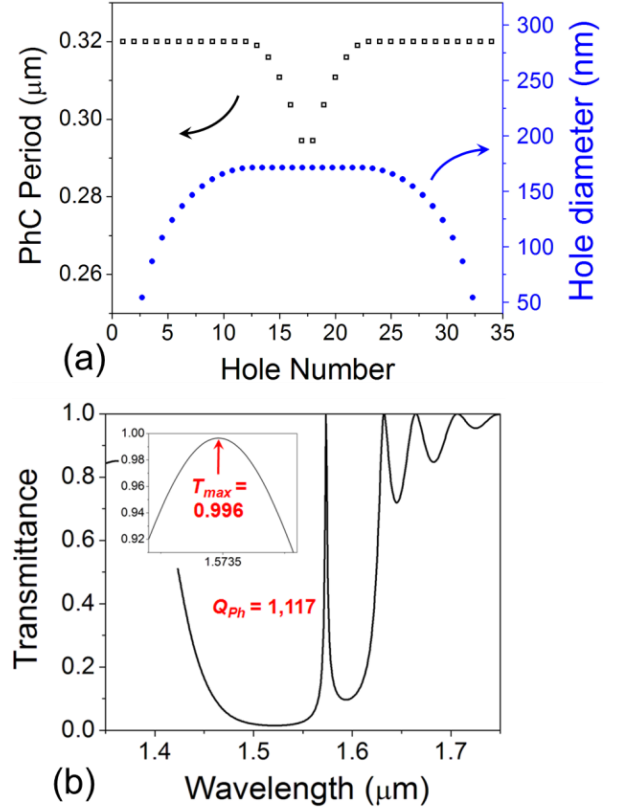

Fig. 4. (a) PhC nanobeam design parameters in the dual cavity modulator structure: the black open squares represent the unit cell period and the blue solid circles label the hole diameters; (b) FDTD simulated optical transmission spectrum of the PhC nanobeam; inset shows the spectrum near the resonance peak at 1.57 μm wavelength.

Performance of the dual cavity modulator can be quantitatively analyzed by combining a coupling matrix method [32] and CMT. We first employ the coupling matrix method to model the static response of the device and extract the basic performance parameters. The detailed coupling matrix formalism is discussed in Appendix III. Subsequently we use CMT to derive the frequency-domain performance of the electro-optically modulated PhC cavity (Appendix IV). This approach is valid provided that the modulator bandwidth is limited by the PhC cavity photon lifetime rather than the ring FSR, a condition we justify in Appendix IV. Based on the coupling matrix method, the EO frequency detuning $\Delta\omega_{EO}$ required to switch between critical coupling and zero coupling is specified by:

$$\Delta\omega_{EO} = \frac{\omega_{Ph}}{Q_{Ph}} \cdot \sqrt{\alpha_{Ph} + \alpha_r} \qquad (12)$$

where $\alpha_{Ph}$, $\omega_{Ph}$, and $Q_{Ph}$ are the insertion loss, resonant frequency and loaded cavity Q-factor of the transmissive PhC cavity, respectively, and $\alpha_r$ is the round trip loss in the micro-ring. This equation, which is valid up to the cut-off frequency of the PhC cavity (~ $\omega_{Ph}/2Q_{Ph}$), dictates the full-swing voltage of the modulator. Since $\omega_{Ph}/Q_{Ph}$ gives the PhC resonant peak FWHM, the resonance shift required for coupling modulation is reduced by a factor of $\sqrt{1/(\alpha_{Ph} + \alpha_r)} = \sqrt{\mathcal{F}/2\pi}$ ($\mathcal{F}$ is the micro-ring finesse) compared to direct intra-cavity index modulation in the PhC cavity. The 3 dB bandwidth of the dual cavity modulator is limited by the smaller value between the micro-ring FSR and the inverse PhC cavity photon lifetime $1/\tau_{Ph} = \omega_{Ph}/2Q_{Ph}$.

Since the round-trip optical loss $\alpha_r$ can be very small in a high-Q resonator and high-speed operation mandates a relatively small $Q_{Ph}$, the main design challenge is to minimize the insertion loss $\alpha_{Ph}$ of the PhC cavity while maintaining a small mode volume. Towards this end, we employ a period-modulated PhC nanobeam cavity design with carefully optimized hole-size-modulated taper structures [33]. Figures 4a and 4b show the PhC design parameters as well as the simulated transmission spectrum. The cavity exhibit a relatively low loaded Q-factor of 1,117 to accommodate fast EO modulation. Top view of the PhC cavity design is shown in Fig. 5a, and Fig. 5b plots the cavity resonant mode intensity profile simulated using 3-D FDTD. The cavity design achieves a loaded Q-factor of 1117 and near-unity (99.6%) transmission according to our FDTD simulations. In our design, we take $\alpha_{ph} = \alpha_r$, which specifies a round-trip loss of 0.017 dB (0.4% power loss). The parameters used in the calculation are listed in Table 1, and details of the simulations are described in Appendix V. Note that we only specify the round-trip loss of the ring: this single parameter is **sufficient** to define the key performance attributes of the modulator device (bandwidth and energy consumption).

In addition, to enhance the optical field confinement and minimize the switching energy, the nanobeam is designed to contain a 20-nm-thick horizontal slot filled with EO polymers sandwiched between two high-index doped silicon strips [34, 35]. The horizontal slot structure can be readily formed by adhesive-free nanomembrane transfer bonding [36] Fig. 5c illustrates the horizontal slot nanobeam waveguide cross-section and its mode profile, which clearly shows the strong field localization in the slot layer. A 4-μm-long section of the PhC nanobeam is electrically contacted from top and bottom through transparent conductor electrodes. When an electric field is applied between the top and bottom contacts, the electric field is highly concentrated in the slot region (Fig. 5f), which maximizes the electric and optical field overlap. To calculate the electro-optically induced PhC cavity resonant frequency shift $\Delta\omega$, we invoke the classical cavity perturbation [37]:

$$\Delta\omega = -\frac{\omega_{Ph}}{2} \cdot \frac{\oint_{slot} \Delta\varepsilon \cdot |E_{Ph}|^2 \, dV}{\oint_{cavity} \varepsilon \cdot |E_{Ph}|^2 \, dV} \quad (13)$$

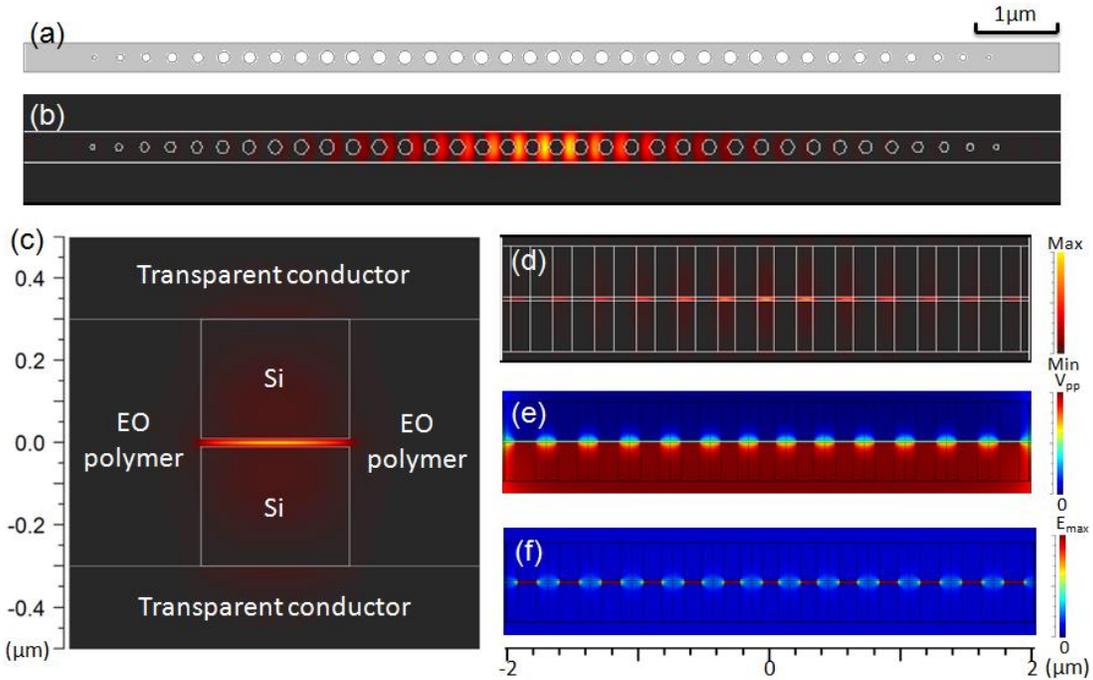

Fig. 5. (a, b) Top view of (a) the period-modulated PhC nanobeam cavity design; and (b) the resonant mode intensity profile simulated using 3-D FDTD; (c) cross-section of the PhC nanobeam along with the nanobeam waveguide mode profile: a horizontal slot is sandwiched between two high-index silicon strips, and a driving voltage is applied through the top and bottom transparent conductor layers; (d, e, f) cross-sectional side view of (d) the optical resonant mode profile; (e) electrostatic potential when an external bias is applied between the top and bottom transparent conductor electrodes of the cavity; and (f) electric field distribution as a result of the applied bias.

Here $E_{Ph}$ is the PhC cavity resonant mode field, and $\Delta\varepsilon$ is the EO polymer dielectric constant modification due to the applied electric field. The integral in the numerator is performed only in the slot region which is filled with the EO polymer, and the integral in the denominator is carried over the entire space. Setting $\Delta\omega = \Delta\omega_{EO}$ by combing Eq. 12 and Eq. 13, we obtain the full-swing modulation voltage of the device to be $V_{pp}$ = 18 mV. This amazingly low driving voltage is a consequence of the efficient coupling modulation scheme as well as the horizontal slot design which optimizes optical and electric field confinement. Capacitance of the PhC cavity is simulated using the finite element method, which yields a low modulator capacitance of 1.6 fF given the small PhC cavity mode volume. Eq. 2 thus points to a record low energy per bit value of 0.26 aJ (i.e. $2.6 \times 10^{-19}$ J). This energy-per-bit figure is more than 1000-fold lower compared to state-of-the-art EO modulator devices. The optical 3 dB bandwidth of the device is limited by the PhC cavity Q = 1117 to be 216 GHz, according to our CMT simulations (Appendix IV). The combination of 0.26 aJ energy per bit and > 200 GHz optical 3 dB bandwidth positions the dual cavity modulator well beyond the performance domains attainable in classical intra-cavity EO modulators, as shown in Fig. 2.

Table 1. Material and device parameters used in the dual cavity modulator design

| Parameter | Value |
|---|---|
| EO coefficient of polymer | 300 pm/V |
| Refractive index of EO polymer | 1.5 |
| Refractive index of transparent conductor | 1.8 |
| Round trip optical loss in the ring | 0.017 dB |

## V. CONCLUSIONS

In this paper, we derive the energy-bandwidth limit in classical intra-cavity electro-optic modulators using a coupled mode theory formalism. The limit prescribes that the minimal energy per bit $W_{lim}$ of classical EO modulators depends quadratically on their optical 3 dB bandwidth as: $W_{lim} = K \cdot f_{3dB}^2$ (Eq. 10), where $K$ is a material constant given by $K = 0.6\varepsilon_r n^{-7} r^{-2}$ fJ/(GHz · pm/V)$^2$ at 1550 nm wavelength.

In the second part of this paper, we demonstrate that this classical limit can be overcome if a coupling modulation scheme is adopted, since coupling modulation speed is not bound by the cavity photon lifetime. By exploiting a novel dual cavity modulator design and strong sub-wavelength optical confinement in nano-slots, the proposed modulator device achieves an ultra-low modulation energy of 0.26 aJ per bit, and a 3 dB optical bandwidth of 216 GHz. This ultra-low energy-per-bit figure represents a dramatic three orders of magnitude improvement compared to state-of-the-art EO modulators, and such outstanding energy-bandwidth performance of the dual cavity modulator is well beyond the classical performance limits, thus making the device an ideal building block for next-generation integrated photonic circuits and optical switching fabrics.

APPENDIX I. CAVITY SHAPE DEPENDENCE OF ENERGY PER BIT

Generally, capacitance of a rectangular cuboid shape cavity can be written as:

$$C = \frac{\Gamma\varepsilon S}{d} \quad (A.1)$$

where $\Gamma$ is a shape-dependent geometric factor, $S$ is the plate area of the capacitor, and $d$ gives the spacing between the two plates. Volume of the cavity $Sd$ can be approximately considered as a constant should the cavity supports only a single resonant mode. The switching energy per bit is then:

$$W = \frac{1}{2}CV_{pp}^2 \sim \frac{\Gamma\varepsilon S}{2d} \cdot (E_{pp}d)^2 = \frac{\Gamma\varepsilon S d E_{pp}^2}{2} \quad (A.2)$$

where $E_{pp}$ denotes the peak electric field required to trigger the EO frequency shift $\Delta\omega_{pp}$. Apparently, the switching energy is independent of the aspect ratio of the cuboid cavity except a geometric factor $\Gamma$ difference of order unity.

APPENDIX II. COUPLED MODE THEORY CALCULATION OF MODULATOR TIME-DOMAIN RESPONSE

The amplitude of an optical resonant mode, denoted as $a$, obeys the following CMT equation:

$$\frac{da}{dt} = i\omega a - \left(\frac{1}{\tau_{in}} + \frac{1}{\tau_{out}}\right) \cdot a + \sqrt{\frac{2}{\tau_{in}}} \cdot S_{in} \quad (A.3)$$

where $S_{in}$ is the amplitude of the input wave, and $\omega$ is the angular frequency of the mode.

First let's consider the time-domain response of a transmissive cavity shown in Fig. 1a to a sinusoidally varying applied voltage given by $V(t) = -\frac{1}{2}V_{pp} \cdot \sin(\Omega t)$. The resonant frequency shift is dictated by the voltage following Eq. 6, which yields the time-dependent resonant frequency:

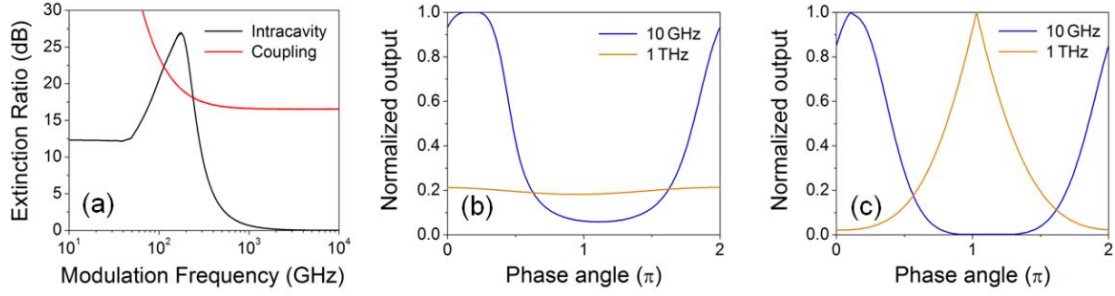

Fig. 6. (a) Frequency-dependent extinction ratios of a resonant modulator whose Q = 1,000. The peak frequency detuning $\Delta\omega_{pp}$ for intra-cavity modulation is twice the resonant peak FWHM; (b) and (c): time-domain response of (b) intra-cavity modulated and (c) coupling modulated devices during one cycle from $t = t_0$ to $t = t_0 + 2\pi/\Omega$. For the coupling modulation scheme, the coupling strength is sinusoidally modulated between 0 and critical coupling.

$$\omega = \omega_0 + \frac{1}{2}\Delta\omega_{pp} \cdot (1 - \sin\Omega t) \quad (A.4)$$

In the equation, $\omega_0$ denotes the time-indpendent angular frequency of the input wave. Assuming a time-dependent complex amplitude $A$ of the resonant mode ($a = A(t) \cdot e^{i\omega_0 t}$) and a fixed frequency input with unit amplitude ($S_{in} = e^{i\omega_0 t}$), the CMT equation A.3 can be analytically solved:

$$A = A_1 \cdot \exp\left(-\frac{2t}{\tau} + \frac{i\Delta\omega_{pp}t}{2} + \frac{i\Delta\omega_{pp}\cos\Omega t}{2\Omega}\right) \\ + \sqrt{\frac{2}{\tau}} \cdot \exp\left(\frac{i\Delta\omega_{pp}t}{2} - \frac{2t}{\tau} + \frac{i\Delta\omega_{pp}\cos\Omega t}{2\Omega}\right) \cdot \int_0^t \exp\left(-\frac{i\Delta\omega_{pp}t'}{2} + \frac{2t'}{\tau} - \frac{i\Delta\omega_{pp}\cos\Omega t'}{2\Omega}\right) \cdot dt' \quad (A.5)$$

where $A_1$ is an arbitrary constant. The first term on the right hand side is the homogeneous solution representing a decaying oscillation in the cavity whose amplitude is determined by the initial condition at $t = 0$. When $t \gg \tau$, this term vanishes. Thus for steady state solutions we only need to consider the second term on the right hand side. The output wave amplitude $S_{out}$ is given by:

$$S_{out} = \sqrt{\frac{2}{\tau_{out}}} \cdot a \quad (A.6)$$

which leads to Eq. 7.

Similarly, the time-domain response of a coupling modulated cavity can also be modeled analytically using CMT if we treat the entire M-Z interferometer as a lumped 4-port coupler element. Note that this assumption is valid only when the modulation frequency is sufficiently smaller than the 3 dB bandwidth of the coupler [38]. An example of the coupling modulated time-domain response is shown in Fig. 6c. Clearly, there is a phase delay between the coupling modulation function and the optical output of the modulator, and the output wave form is significantly distorted as the modulation frequency increases. Both factors contribute to a nonlinear response of the coupling modulation scheme, which can be compensated by choosing an appropriate input wave form in practical applications.

Figure 6a compares the extinction ratios as functions of the modulation frequency when the two modulation schemes are implemented in a resonant modulator whose Q = 1,000. Unlike intra-cavity modulation, coupling modulation is not bound by the cavity photon lifetime and thus no significant roll-off at the high frequency end is observed.

APPENDIX III. COUPLING MATRIX ANALYSIS OF THE DUAL CAVITY MODULATOR

Taking the input wave amplitude as unity, we label the complex amplitudes of waves at different ports of the coupler as is shown in Fig. 3b. The coupling matrix of the lumped coupler can be obtained by cascaded the coupling matrices of the two lossless 3 dB couplers and the MZ interferometer (with power balanced arms):

$$K_c = \begin{bmatrix} t & k \\ -k & t \end{bmatrix} \cdot \begin{bmatrix} \sqrt{1-\alpha_{Ph}} & 0 \\ 0 & \sqrt{1-\alpha_{Ph}} \cdot e^{i(\pi+\varphi)} \end{bmatrix} \cdot \begin{bmatrix} t & k \\ -k & t \end{bmatrix} = \sqrt{1-\alpha_{Ph}} \cdot \begin{bmatrix} t^2 + k^2 \cdot e^{i\varphi} & tk - tk \cdot e^{i\varphi} \\ -tk + tk \cdot e^{i\varphi} & -k^2 - t^2 \cdot e^{i\varphi} \end{bmatrix}$$

$$\sim \sqrt{1-\alpha_{Ph}} \cdot \begin{bmatrix} 1 + k^2 \cdot i\varphi & -tk \cdot i\varphi \\ tk \cdot i\varphi & -1 - t^2 \cdot i\varphi \end{bmatrix} = \sqrt{1-\alpha_{Ph}} \cdot \begin{bmatrix} 1 + \frac{1}{2}i\varphi & -\frac{1}{2}i\varphi \\ \frac{1}{2}i\varphi & -1 - \frac{1}{2}i\varphi \end{bmatrix}$$
(A.7)

where $t = k = 1/\sqrt{2}$, $\alpha_{Ph}$ and $\varphi$ are insertion loss and electro-optically induced phase delay through the PhC cavity, respectively. If the micro-ring cavity has a high Q-factor, the phase delay $\varphi$ needed for critical coupling is small ($\varphi \ll \pi$), and we Taylor expand the term $e^{i\varphi}$ in the equation to obtain the final result. The critical coupling condition specifies:

$$1 - \frac{1}{4}\varphi^2 = (1-\alpha_{Ph}) \cdot (1-\alpha_r) \sim 1 - \alpha_{Ph} - \alpha_r \quad (A.8)$$

The phase shift $\varphi$ may be expressed in terms of PhC cavity's group delay $\tau_g$ and its EO frequency shift $\Delta\omega_{EO}$ as:

$$\varphi = |\tau_g \cdot \Delta\omega_{EO}| = \frac{2Q_{Ph}}{\omega_{Ph}} \cdot \Delta\omega_{EO} \quad (A.9)$$

Combining Eqs. A.8 and A.9 leads to Eq. 12

APPENDIX IV. FREQUENCY-DOMAIN RESPONSE CALCULATION OF THE DUAL CAVITY MODULATOR

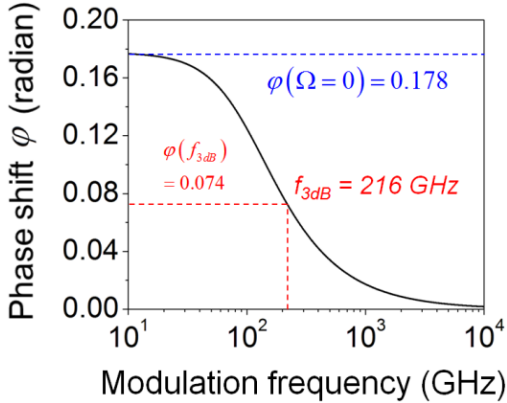

Fig. 7. Phase delay in the EO-modulated PhC cavity as a function of modulation frequency calculated using the coupled mode theory: the static response $\varphi(\Omega = 0)$ corresponds to the phase delay required to switch the micro-ring from the "off" state (zero coupling) to the "on" state (critical coupling). At 216 GHz (the PhC cavity photon lifetime limited bandwidth), the micro-ring "on" state extinction ratio rolls off to 3 dB.

Frequency-domain response of the modulator was simulated using the CMT framework presented in Appendix II. The modulation bandwidth of the device is limited by two factors: the inverse PhC cavity photon lifetime and the micro-ring FSR. Figure 7 plots the phase delay $\varphi$ induced by electro-optic resonance shift of the PhC cavity as a function of modulation bandwidth. At zero frequency, the phase delay required to reach critical coupling is calculated using Eq. 12 to be $\varphi(\Omega = 0) = 0.178$ radian. The optical 3 dB bandwidth of the device corresponds to a phase delay $\varphi$ that results in 3 dB extinction ratio of the ring resonator in its "on" state. Since $\varphi$ defines the amplitude coupling coefficient into the micro-ring (Eq. A.7), it is straightforward to derive from the coupling matrix method that $\varphi(f_{3dB}) = (\sqrt{2} - 1) \times \varphi(\Omega = 0) = 0.074$ radian. Therefore, we can read from Fig. 7 that the PhC cavity photon lifetime limited bandwidth of the dual cavity device is 216 GHz.

Next we argue that the micro-ring FSR does not pose a bandwidth constraint on our design, as the ring FSR limited bandwidth can be much larger than the PhC cavity photon lifetime limited bandwidth. To operate in the photon lifetime limited regime, the ring FSR has to be greater than 216 GHz, corresponding to a silicon micro-ring radius of ~55 μm or smaller. The lower limit of the ring radius is posed by radiative bending loss; for silicon micro-rings, bending radius as small as 10 μm still provides negligible bending loss given their high index contrast and tight optical confinement [39]. As a specific example of the micro-ring design, if we choose a ring radius of 20 μm, the round-trip loss of 0.017 dB ($\alpha_{ph} = \alpha_r = 0.004$) then specifies a micro-ring waveguide loss of 1.35 dB/cm, which is well within the reach of state-of-the-art nanofabrication technology [40].

APPENDIX V. FDTD AND FEM SIMULATION METHODOLOGY

3-D FDTD simulations are performed using the commercial package of RSoft FullWAVE$^{TM}$. All PhC cavity simulation results (resonant wavelength, cavity Q, and insertion loss) presented in this paper have passed our convergence test with a relative error of less than 1%.

The device capacitance and electric field distribution (Fig. 5e and 5f) are simulated using a 3-D finite element method using the commercial package of COMSOL Multiphysics$^{TM}$. The simulated electric field distribution is used to calculate the electro-optically induced polymer dielectric constant change $\Delta\varepsilon$ in Eq. 13.